\documentclass{elsart}

\usepackage{natbib}
\usepackage{graphics}
\usepackage{epsfig}

\begin{document}
\runauthor{Walton \& Smartt}
\begin{frontmatter}

\title{The ING Instrumentation Conference Discussion \\
{\small Options for a Competitive Observatory }}

\author[ING]{N.A. Walton \& S.J. Smartt\thanksref{IOA}}

\address[ING]{Isaac Newton Group of Telescopes, Apartado 321, 
38700 Santa Cruz de La Palma, The Canary Islands, Spain}

\thanks[IOA]{Present Address: Institute
of Astronomy, University of Cambridge, Madingley Road, CB3 0EZ, UK}

\begin{abstract}

An expert panel initiated discussion on a number of key questions
facing the role of 4-m and small telescopes in the new era of 8-m
telescopes. The panel and audience agreed that the 4-m class telescope
role would necessarily evolve, but would still be important in the
coming years. The need for an active development programme of
competitive instrumentation for 4-m class telescopes, and in
particular the William Herschel Telescope (WHT) was stressed. In
conjunction with this, the need to de-commission instrumentation made
redundant by 8-m class telescopes was noted. New operational modes,
including greater emphasis on survey programmes, and possibly queue
scheduling, coupled with changes to the procedures for allocating time
were seen as desirable. The panel and audience supported the Isaac
Newton Group's emphasis on the development of instrumentation to
exploit its imminent deployment of the WHT's facility Adaptive Optics
system.

\end{abstract}


\end{frontmatter}

\section{Introduction}

In order to gauge the views of the Isaac Newton Group's
(ING)\footnote{http://www.ing.iac.es} user community, a general
discussion session was organised on the afternoon of Thursday, 8th
April 1999. The discussions covered a range of topics, and provide
some interesting insights on issues currently facing the ING. These
may be of use when considering the development of other sub-8-m
telescope facilities in the coming decade. The discussions were
chaired by Prof.\ Phil Charles from the University of Oxford. An
expert panel (see Table~\ref{tab-1}) was assembled who initially
responded to each of the discussion points before inviting input from
the conference attendees.

For ease of editing, and accuracy, the discussion session was recorded
on audio tape. The editors have transcribed the audio tapes, and
edited the text in particular drawing out important arguments and
pruning repetition and side issues raised.  The discussions flowed in
a fairly logical fashion, and the editors have added discussion point
titles at the start of each new discussion section.  However there are
some inevitable transgressions and asides.

The editors take full responsibility for this text, and apologise if
they may have in any way mis-interpreted or mis-understood any of the
contributors comments. 

\begin{table*}[ht]
\caption{Members of the discussion panel and affiliations} 
\label{tab-1}
\begin{center}
\begin{tabular}{ll} \hline
Name & Institution \\ \hline 
Phil Charles (Chair) & Oxford \\
Mike Edmunds & Cardiff \\
Jim Emerson & Queen Mary \& Westfield, London \\
Roger Davies & Durham \\
Pat Roche & Oxford \& UK Gemini Office \\
Ren\'{e} Rutten & Director, Isaac Newton Group \\
\hline\end{tabular}
\end{center}
\end{table*}

\section{The Discussion Session}

\subsection{Opening Comments}

{\em Phil Charles:} I am standing in for Carlos Frenk who
unfortunately could not attend this meeting.  We had agreed several
things with Carlos and that was the original initial list of topics up
there (see Table~\ref{items}). The panel is here to help keep
discussion going.  It's your opportunity to take part in this
discussion as well.

In preparing for this, we came up with this list of questions.  We
have decided to drop the first one since we could go on for a week
without ever getting an agreement on it. And obviously we want to be
aware of the underlying science, but we want to concentrate here on
the instrumentation and the future direction of our facilities.  You
are the users and this is your opportunity to provide input into this
whole process.

\begin{table*}[ht]
\caption{Questions posed to the Expert Panel during the discussion session}
\label{items}
\begin{center}
\begin{tabular}{l} 
What are the current major astronomy science issues? \\ 
What instrumentation is needed to address these issues? \\
What science is best carried out on 4-m or 2-m class telescopes? \\
What facilities might now be redundant? \\
La Palma and GranTeCan, the Spanish 10-m telescope. \\
What role for `robotic' telescopes? \\
Which telescope/instrument operational modes should be developed? \\
The position of `survey' programmes? \\
Opportunities for `networking' European sub 8-m telescopes \\
\end{tabular}
\end{center}
\end{table*}

{\bf Operational Modes and Scheduling the ING and other UK telescopes:
the Impact of Gemini.}

{\em Phil Charles:} With Gemini [{\em two 8-m telescopes constructed
by a multinational consortium including the
UK}]\footnote{http://www.gemini.edu} coming along and hopefully
GranTeCan [{\em The Spanish 10-m telescope currently in the early
stages of construction}]\footnote{http://www.gtc.iac.es} not too far
afterwards, this will have an impact on the way we operate and the way
we allocate time.  The first question is about the PATT [{\em the UK's
Panel for the Application of Telescope Time}] mode, the way we propose
for telescope time, the role of the separate time allocation
committees [{\em TACs}]. We heard this morning that we have a lot of
TACs handling our telescopes. And yet space based facilities allocate
time according to strategic priorities. At the ING, the Joint Steering
Committee [{\em the management board of the ING, and now
re-constituted as the ING Board}] have decided to target a certain
fraction of the 2.5-m Isaac Newton Telescope [{\em INT}] time for
survey work. Is that the direction we should be going?  Is it enough?
Is it too much?

{\em Jim Emerson:} At this meeting, people from the NOAO [{\em The
USA's National Optical Astronomical
Observatories}]\footnote{http://www.noao.edu} and Dr.\ Roland Bacon
from Lyon have told us that both at the NOAO and in France time was
allocated via thematic TACs.  Danny Lennon talked about the
difficulties of assigning time and the disparity between getting
either three hours via an ING service programme or a week from a PATT
application.

Pat Roche\cite{roche} mentioned Gemini programmes; a science project
may need a combination of imaging and spectroscopy, perhaps utilizing
instruments on smaller telescopes as well. At present this would need
multiple applications to the varying TACs.  I would like to propose
that PPARC [{\em the UK's Particle Physics and Astronomy Research
Council}]\footnote{http://www.pparc.ac.uk} appoint a small working
party to look at the possibility of restructuring PATT to allocate
time to specific projects, allocating time on whichever facilities are
needed.  I can see some difficulties but it sounds like there would be
many advantages.

{\em Phil Charles:} I think it was Richard [Bower] who asked about
the problem that you can not get more than 3 hours in service
mode. Danny [Lennon] mentioned the expense of running queue [{\em Q}]
scheduling which you could describe as an enhancement of service.  The
additional cost of the queue mode is one I recall from my time on La
Palma --- if you do use people then it may be more expensive because
you have to pay for them to be there, lets say, you used two or three
nights a month for this type of work, so 30, 35, 40 nights a year.
But that's 40 nights of time for which the observers have not had to
travel out.

But this is money saved from a non-observatory budget to which the ING
has no access. If Q scheduling is activated in a significant fashion,
the observatory needs to be compensated with the savings in travel \&
subsistence. Roger, does that agree with what has been agreed or
discussed in Gemini?

{\em Roger Davies:} This is not only possible but it has been done
already.  2dF [{\em a multi-object fibre spectrograph}] on the Anglo
Australian Telescope [{\em AAT}]\footnote{http://www.aao.gov.au} has
an observer who is paid for by PATT --- you can't go and observe with
2dF even if you want to.  If people wanted to do this it could be
decided and done I think. Would you agree Colin?

{\em Colin Vincent:} That's correct. It is also done for SCUBA on the
JCMT [{\em James Clark Maxwell
Telescope}]\footnote{http://www.jach.hawaii.edu}.

{\em Vik Dhillon:} If thematic panels are instigated for various
subject areas, how do you decide how much time to give each subject?

{\em Jim Emerson:} That would be hard to decide. Initially you could
go through the last few rounds of PATT and divide it up in the same
proportions.  But clearly the problem would be that you would then
have philosophical arguments over whether cosmology was more important
than stars before you got to the thing. You could have some rolling
average. The working group should explore the approach taken by the
Americans and the French.

{\em Bruce Bohannan:} Go and look at HST [{\em Hubble Space
Telescope}]\footnote{http://www.stsci.edu}, go and look at what NOAO
is doing, starting this semester. Look at what the French do, and take
what works best.

{\em Phil Charles:} Yes, the HST takes a strategic approach.  The
panels allocate certain fractions of time between different scientific
areas. Now I would actually say that has many strong points going for
it. And it would prevent a small TAC ignoring particular scientific
areas, perhaps because that TAC has a group of people who all think,
CVs [{\em Cataclysmic Variables}] for instance are important, and thus
give all the time to CVs.

{\em Tim Hawarden:} Colin mentioned that SCUBA on the JCMT has been
operating about three different forms of high intensity Q scheduling
for some time. Of which the Dutch system seems to me to be working
best --- here the mode is that visiting astronomers plus resident
astronomers do the work. Everyone goes there and does service mode
observing and sometimes it's your work and sometimes it's somebody
else's work.

{\em Clive Tadhunter:} If allocation is done by dividing it up by
subject, then to make it efficient you would have to do it for all the
telescopes. Ground based telescopes represent a more complex
scheduling situation than for instance the HST. They have many
instrument configurations, and there are weather variables to
consider.  I find it very difficult to see how you would get a
workable Q system with so many instruments, so many different colours
of the night, etc.

{\em Peter Sarre:} The discussions seem to have been a question of all
or nothing.  Some of the research councils have a blend of themes and
response mode.  I wonder if there are themes which can be time
dependent in their size and vary after a couple of years when they are
reviewed, but then half the time is also freely available for the
innovative programmes that don't naturally fall into those modes.

{\em Konrad Kuijken:} ESO [{\em European Southern
Observatory}]\footnote{http://www.eso.org} have been running seven to
eight telescopes in the last years with six or so subject panels.
These all get together, the different panels give away surplus time on
some telescope they didn't need on that particular round and they buy
other time with it. In the end it seems to work reasonably well.

{\em Phil Charles:} Also the panels handle more than one telescope.

{\em Konrad Kuijken:} The panels cover all the telescopes including
the VLT [{\em ESO's Very Large
Telescope}]\footnote{http://www.eso.org/vlt}.  What happened at the
last VLT round was that, for instance, the planetary scientists got
three nights of VLT time to divide among them in proportion to the
amount of time that was solicited.  And it worked on a historical
basis, based on a rolling average.

{\em Phil Charles:} Would the majority of people here like to see
significant changes to the PATT allocation method? Who thinks that
moving to this kind of thematic approach is a good idea? Who thinks
the present PATT mode is working well?

{\em [Ed's note: on a show of hands $\sim$70\% of the attendees agreed
that the present PATT allocation system should be revised in the light
of the arrival of the Gemini telescopes.]}

{\bf The position of survey programmes.}
 
{\em Richard Bower:} Isn't that a bit of an unfair choice? I don't
see how having thematic panels would help with this decision between
putting forward surveys and response mode. Because the surveys are
very often multi-disciplinary and thus would fall between different
thematic panels.

{\em Phil Charles:} I saw `surveys' as being one of the themes.

{\em Roger Davies:} I think they are unrelated issues.

{\em Pat Roche:} The fundamental question is, are we putting ourselves
at a competitive advantage or disadvantage.  The way we operate now
sets up the telescopes in competition rather than a collaboration.  I
believe that this is bad and we are going to pay very dearly if we
continue to operate in that mode.  We can help to try and overcome
this is by blurring the boundaries between telescopes, encouraging the
telescopes to operate corporately in order to facilitate
significant scientific progress.

Further, it is clear that the number of programmes which are done on a
multi telescope basis is going to increase and therefore allocating
small packets of time on individual telescopes will make less and less
sense. This is inevitable and I believe that we risk putting ourselves
at a significant disadvantage unless we change the way time is
allocated across telescopes. 

{\em Andy Longmore:} I agree with Pat.  We have seen examples of this
in the past. Years ago when surveys of high red shift galaxies were
popular, the University of Hawaii [{\em UH}] dedicated significant
UKIRT [{\em United Kingdom Infra Red Telescope}] time to this
project. UK groups used to complain that they could not get enough
time to compete with the UH block allocation. Thus, at a time when the
UK were leading in a field, we couldn't compete with our own
telescope.

I think perhaps the advent of the of the 8-m telescope is really
bringing this to a head. This forum here enables people to say what
they think the pros and cons of the various time allocation options
are. It is essential to decide on how to take these discussions
forward.

{\em Phil Charles:} Ten years ago I had a conversation with the
Chairman of PATT. The handful of top proposals were obvious. The very
bad proposals were even fewer.  The committee then spent the majority
of its time trying to separate out the 2.3 versus the 2.4 versus the
2.5 average proposals.  The Chairman suggested that one could randomly
decide on these by drawing from a hat, and have no impact on the
quality of the science being allocated.  I don't know if the Chairman
of PATT of the time wants to say anything?

{\em Mike Edmunds:} You summarised that very well.  There is no ideal
system.  The new problem we are facing is really co-ordination between
allocation of time. That's the critical issue, not whether it is one
particular subject or whatever, it's how you co-ordinate between
awarding time on different facilities.  You've just got to find a way
that does allow you to co-ordinate well.

{\bf Thematic Time Allocation Committees.}

{\em Roger Davies:} I don't really agree with many of the things Mike
said in the sense that I think that the division now is a facility
based division which has historical roots.  When I started my career,
PATT was only the AAT and the INT.  Telescopes have come on-line one
at a time, each added to PATT with its own separate panel TAC.

I've done a lot of programmes that need, for example, optical
photometry, infrared photometry, and optical spectroscopy. It takes a
lot of PATT rounds to get one project finished that requires those
simple observations on many telescopes. 

We are now in a situation where we're going to have our competitors
coming forward to the Gemini international TAC and they will be able
to say `we've already allocated time on the Kitt Peak 4-m, on
WIYN\footnote{http://www.noao.edu/wiyn/}, on
CTIO\footnote{http://www.ctio.noao.edu}, for this programme and we now
need the infrared spectroscopy with Gemini'. They will get the time
because they've got the backing and they will get the programme
finished in one shot.

The TACs need to service the science programmes, and not the
telescopes.  From the point of view of competitiveness, and getting
the science done and into the journals, we'd be better off allocating
the time for a scientific programme on the telescopes it needs, in
more or less one go.  The way we think about this problem need to
change. We need to alter the way in which we apply for telescope time
so that you apply to solve a scientific problem and not to get nights
on a 3-m telescope.

{\em Jim Emerson:} Can I re-iterate my suggestion at the beginning
that this meeting suggests that PATT convene a working group because
we are not going to solve this here. Does the meeting feel that is a
sensible thing to do, as long as they weren't on the panel of course.

{\em Pat Roche:} It would be helpful to know if most people feel this
is an issue or not.

{\em Mike Edmunds:} I think is very important that there is the option
of keeping the status quo.  You've got to think very carefully before
implementing a new allocation system, to ensure that it really does
work as you want.

{\em Phil Charles:} I think both Pat and Roger have outlined that there
are significant problems with the current system which are: we get by
at the moment but we are going to be at a severe disadvantage as
we move into the Gemini and then GranTeCan area.

{\em Reynier Peletier:} So what about the other telescopes?  What about
UKIRT, Gemini, in the future. If you combine things do you want to
limit yourselves to one site?

{\em Mike Edmunds:} No, This would include AAT and UKIRT and Gemini and
everything.

{\em Richard Bower:} What about weather. Roger's point was a very
valid one --- you would want to be allocated the time to complete the
project within one semester.  But you often loose so many nights due
to the weather that you need to come back for another semester.  So if
we are going to revise the way the telescopes are allocated are we
going to revise the way that weather is taken into account.

{\em Pat Roche:} That is already happening of course at UKIRT and some
other optical telescopes, the highest rated programmes are started and
there is a reactive schedule programme.

{\em Roger Davies:} To correct what you said Richard, I didn't say
that people should do their programmes in one semester, rather they
need to plan them better. I wouldn't suggest it is sensible to try and
complete a programme in one semester because you learn as you go
along. The first semester can be used to confirm the method, before
significant progress in the second and following semesters.

You need applicants to think about how they are going to implement
their programme scientifically rather than writing a telescope time
application to justify `x' nights of time.  I think that we should
just be presenting the science programme and trying to get the science
done and trying to take away a little bit of the lottery effect in
time allocation.

{\em Colin Vincent:} It seems like the consensus of the meeting is that
you would like see this taken forward so I would be happy to take it
back to PPARC and suggest to them that some sort of group is set up to
consider it further.

{\bf Operational Modes: Q scheduling.}

{\em Phil Charles:} What would the ING need or how they would they
change to create the infrastructure to handle significant amounts of Q
scheduling.

{\em Reynier Peletier:} At some telescopes, instead of applying for
nights, you apply for objects, as for the HST. With Q scheduling, the
observatory could guarantee that your object gets done.  Basically
twice the amount of time is scheduled for the amount.

{\em Phil Charles:} With the Q scheduling you don't have to allocate
twice the amount of time. High priority programmes get done during the
first clear night(s).

{\em Reynier Peletier:} No, I am saying they guarantee it is done, 
not with a priority because then you may not get your data.

{\bf Instrumentation Rationalisation at the ING.}

{\em Phil Charles:} Do we have too many instruments on the WHT [{\em
William Herschel Telescope}] and should we retire LDSS-2 [{\em Low
Dispersion Survey Spectrograph}] and TAURUS-2 [{\em an imaging
fabry-perot spectrograph}]? Does LDSS-2 provide particular
capabilities that WYFFOS [{\em a fibre fed spectrograph at the WHT}]
doesn't.  We heard about the developments for the future, meaning more
instrumentation. Yet the ING are going to be limited in the number of
instrument changes that they are allow because of the cost associated
with that. Hence rationalisation of the ING instrument suite may be
needed.

{\em Andy Longmore:} If we are making the case that the science for
these instruments is really good and there is an excellent niche that
is worth supporting, are we giving up a bit too soon. If the science
is there and it may justify more staff resource to support the extra
instruments. 

{\em Ren\'{e} Rutten:} My fear is that a year from now, we will have NAOMI
[{\em the facility Adaptive Optics (AO) system for the
ING}]\footnote{http://www.ing.iac.es/$\sim$crb/wht/ao.html} which
potentially could be a major drain on ING resources. We can't wait for
another year to make a decision on instrument rationalisation, because
the de-commissioning process can take two years.  Something has to
happen now in order to free the resources of the ING to enable
adequate resource to be available to support upcoming new
instrumentation.

{\em Roger Davies:} I would like to thank Ren\'{e} and his staff for the
support they gave to SAURON [{\em an Integral Field
Spectrograph}]\footnote{http://www-obs.univ-lyon1.fr/$\sim$ycopin/sauron.html}
when it was commissioned on the WHT. The team on the site did an
outstanding job. SAURON was scheduled in a single observing block.
However during the run there was a problem with the A\&G [{\em
Acquisition \& Guidance unit}] and SAURON needed to be removed.  The
team were outstanding, replacing a broken bearing in one day, and
having SAURON aligned and ready for the evening.  This illustrates
that you are not always in control of the level of support that you
might need.  We always have to deal with the inevitable support
problems like those I have just described.  Thus, it seems restricting
to be discussing de-commissioning instruments for the sake of the few
extra days of effort needed to put them on and off the telescope.

I would like to speak about LDSS-2. We are not going to get time on
10-m and 8-m telescopes to take spectra of 21st and 22nd magnitude
objects.  It's not competitive.  We do that with 4-m telescopes.  And
we have to have spectrographs that are effective at those quite faint
levels.

I think the AAO have shown what can be done if you put a little bit of
investment in an old instrument. By replacing the CCD and implementing
charge shuffling, they increased the efficiency of LDSS-1 [{\em LDSS
on the AAT}] which is much less automated that LDSS-2 by a factor of
2--3 [{\em the upgraded instrument is named LDSS++}]. The charge
shuffling allows the number of object that can be obtained
simultaneously to be increased from a few tens to a few hundreds, an
astonishing efficiency game.  This was done for a few tens of
thousands of dollars and some of Karl Glazebrook's and a few others
time.

What I would like to see is for the ING to apply a similar
transformation to the LDSS-2 --- we make a modest investment to bring
it back up to competitive standards.  Mask cutting for LDSS-2 is
proving difficult and it is not quite as accurate as it used to
be. Perhaps one can learn from the systems that will be put in place
for GMOS [{\em Gemini Multi-Object Spectrograph}] and other mask
cutting spectrographs to learn how to do that. I don't think there
needs to be a big investment in LDSS-2, but it does need to be kept on
the telescope and supported. I think if you loose that you have lost
your faint light spectroscopy. The 4.2-m WHT is still a big telescope.

{\em Ren\'{e} Rutten:} To provoke the discussion a bit further with a bold
statement, which I do not necessarily agree with personally.  There is
a wonderful LDSS and TAURUS on the AAT, both more capable currently
than the similar systems on the WHT. Thus, let's do that work at the
AAT, and forget about it at the WHT.

{\em Richard Bower:} The advantage we have in the northern hemisphere
is the INT wide field camera.  The role of the 4-m telescopes is
following up spectroscopically the objects that are found on the INT
using the wide field camera. Those objects are going to be potentially
faint and are going to be over a wide field of view.  LDSS-2 has some
very strong advantages in following those objects up, in particular if
you want to look at distant clusters. WYFFOS has some other advantages
but you can't get the close packing of the objects that you can on
LDSS-2. So we need to think strategically on what is the right way to go
to provide the spectroscopic instrumentation for the next decade.

{\em Paul Groot:} From the discussion this morning and yesterday, the
feeling appears to be there should only be one instrument on the WHT
for faint object spectroscopy.  I would like to put a pro-WYFFOS
statement.  WYFFOS is a flexible system in the sense that it can be
fed from different instruments. It may be possible to upgrade WYFFOS
to give the equivalent performance of LDSS++.

Are there too many instruments? I know from experience that the INT
wide field camera is now working very well. Why not de-commission prime
focus imaging at the William Herschel and have only prime focus
spectroscopy there.

{\em Nic Walton:} For the ING instrumentation, developments can be put
forward at differing resource levels. Various low cost incremental
upgrades can be made in the AF2/WYFFOS area [{\em AF2 is the WHT's
fibre feed unit to WYFFOS}]. The packing issue can be addressed by
employing IFUs. At another level TAURUS-2 with a TTF could be combined
with LDSS to deliver increased functionality --- along the lines
suggested by Joss Bland-Hawthorn at the AAO, this might be a proposal
to the GBFC.

{\em Richard Bower:} I agree with Nic. What is required for LDSS is
just a collimated space in which you put the grism. That fits in very
well with improvements to TAURUS-2. Combining the two instruments is an
extremely efficient way to go. You can even use it for polarimetry.

{\em Mike Edmunds:} If you have a limited budget with which to operate
and instrument a telescope, then you have to prioritise which
instruments to support and/or develop. What are the most important
instruments, what will do most of the science we want to do at the
present time?  How much can we afford? How efficiently can we keep
these instruments running and up to specification and reliability. It
is much better to have three instruments that work extremely well than
ten or more, poorly supported, instruments.

{\em Phil Charles:} If you are going to demand that the science for
all the instruments is essential and that you do need the manpower to
maintain them and operate them with that frequency, then we may wish
to divert money from building new instruments towards maintaining
those that we have. 

{\em Clive Tadhunter:} It is not just manpower, its also {\em
scientific efficiency} because every time you change an instrument you
need an extra night to set it up. There is a limit as to how many runs
you can have, and I think we have reached that limit on the WHT.

{\em Phil Charles:} Frequent changes also impact on the reliability of
the instrumentation.

{\em Tim Hawarden:} We've run on UKIRT with a minimum number of
instruments principally for historical reasons. We too, in an era of
shrinking budgets, now have to accommodate more instruments than we
have right now. But we do normally have more than one instrument on
the telescope, where these instruments remain ready for use for long
periods of time. We work with instruments that have to be extremely
reliable. They stay cold and ready for use for months and months on
end, and perhaps there is a way forward here for simplification. If
you have got something like eight instruments which are moved on and
off the telescope at regular intervals and need alignment every time
and adjustment every time they go on and off, perhaps a new
fundamental approach to placing instruments on the telescopes is
needed.

{\em Ren\'{e} Rutten:} To explain the situation.  The problem concentrates
on the Cassegrain focus where there is one very popular instrument and
a number of less popular instruments and a number of visitor
instruments.  So it is that focus which is particularly pressured and
complicated regarding instrument changes.

{\em Phil Charles:} This discussion has so far focussed on the need to
reduce instruments in order to accommodate NAOMI and the other extra
instruments that are coming on line next year. But in the longer term, say
four years, UK involvement with GranTeCan will force us to make very
serious choices on how to operate the WHT.

{\em Roger Davies:} Mike [Edmunds] outlined a clear scheme for
prioritising the use of instruments, emphasising high reliability by
reducing the number of times you change instruments.  But nobody
mentioned doing world class science.  It is relatively easy to devise
a way of running the observatory to stay within a budget.  The {\em
big problem is that we are not competitive if we do that without
maintaining world class instrumental capabilities}.  I think that only
after prioritising the science, can you make the trade-off as to which
instruments/ capabilities are needed.

I would be astonished if instruments with a ten arc minute field of
view and bigger on 4-m telescopes working at the sky background limit
are not competitive into the 8-m age. This will still be
competitive. Not many of the 8-m telescopes have large field of views,
and those that do have as yet have limited instrumentation there. And
it's inefficient to use those 8-m telescopes to take 21st and 22nd
magnitude spectra.

{\em Mike Edmunds:} I wasn't suggesting that we don't do world class
science. I didn't say which three instruments or for what.  One of
those could be extremely innovative and extremely new. You fully
exploit it, get it working well until you've used it then put
something else on as your next instrument development.  I do wonder
whether we really should be looking more towards programme
orientation. We have one general purpose spectrograph, an imager and
something completely novel.  You can't do all science though.

{\em Clive Tadhunter:} Yes but we are nowhere near three instruments
yet on the WHT. We've got something like ten to thirteen.  And I think
that's a bit too much.

{\em Pat Roche:} There is also the parallel with the AAT. There is 2dF
\& WYFFOS, the two TAURUSs and the two LDSSs.  Duplicated instruments, 
with the last two not getting a lot of time at the moment.  In general
these instruments don't care where they look as they support
cosmological wok. 

{\em Roger Davies:} That's absolutely not true. If you try and find
clusters of galaxies whose red-shifts put absorption lines between the
night sky lines long wards of 700 nanometers, there are not that many
clusters.  I think it's a great strength of our UK programme that we
have instruments that are duplicated in both hemispheres. It means you
can do the same experiment in both hemispheres, you can do surveys
with similar instrumentation. That's one of the great strengths of
Gemini.  I think it would be a pity if the AAT or the WHT diverge in
that capability.

The AAT solution to the same problem is the proposal of the ATLAS
spectrograph \cite{bailey} It combines a replacement in the WHT
context for ISIS [{\em the long slit spectrograph at the WHT}], with
LDSS and TAURUS, and increases the field to 24 arcmin. The AAT Board
is very interested in this, and perhaps could be an instrument for
development at the WHT as well.

{\em Ren\'{e} Rutten:} The discussion has so far focused on LDSS-2.  I also
mentioned TAURUS-2 as having some uncertainty regarding its future.
Could the TAURUS-2 users say something about what they see as the long
term future of TAURUS-2.

{\em Clive Tadhunter:} I would just like to echo what Nic [Walton]
said, that TAURUS-2 needs to be upgraded if it is to be effective in
the future.  Perhaps combining it with LDSS is the thing to do.

{\em Johan Knapen:} In my talk tomorrow\cite{knapen} I will give some
advantages of TAURUS-2. It is ground breaking in certain areas where
it's really unique for the UK at the moment.

{\bf Longer Term Instrumentation Development: addressing major
astronomical science with sub 8-m telescopes.}

{\em Phil Charles:} I should remind you that we would also like
peoples input into the longer term ING instrumentation development. 

{\em Ren\'{e} Rutten:} People asked me about the percentage of time used
for the different instruments and I have a table of statistics here,
slightly out of date covering the two years, 1996/97.

WHIRCAM [{\em the near infra-red camera for the WHT}] was used about
8\% of the time, with its replacement, INGRID [{\em the new near
infra-red camera for the WHT}] that will go up substantially. AF2 is
only 6\% because it was then being commissioned: for 1998 and 1999 its
use is much higher.  The more established instruments are
approximately stable in their percentage use.  ISIS is the workhorse
instrument, prime focus is used $\sim$10\% of the time, LDSS-2 and
TAURUS-2 lie a little below that.

Concerning LDSS and TAURUS-2.  We can't have everything all the time, in
the North and the South, and at no cost.  We need to understand our
priorities.  Both these instruments have opto-mechanical problems and
need serious overhauls. This can't be done in the short term with out
delaying the commissioning of INGRID with the AO system, and the
development work on AF2/WYFFOS. These are the priorities we are
looking at.  Is it worth it --- yes or no?

There is also an impact on the development programme. I would like to
go away from this meeting with an indication from the community
whether the {\em priorities} for the development programme, {\em
adaptive optics (focusing on the optical) and wide field multi-object
spectroscopy (optical and the non-thermal near IR)}, are correct for
the mid term future.  I need to know what the community wants us to
bid for within GBFC and NWO [{\em UK and NL}] funding routes.  There
is money earmarked for a major new WHT instrument, if we don't have a
proposal for it, that finance will be lost to the ING.

{\em Is imaging on the WHT at the prime focus important?} I made an
explicit statement \cite{rutten} that the current development plan for
the WHT envisages no enhancement to the prime focus imaging
capability.  Some other 4-m telescopes (e.g. CFHT, [{\em Canada France
Hawaii Telescope}]\footnote{http://www.cfht.hawaii.edu}) are pursuing
these developments rigorously and therefore it may be too late for the
WHT to catch up in this area.

Finally, we have a re-vamped instrumentation working group. We had the
first meeting of the new instrumentation working group two days ago [6
April 1999].  I would like to invite everybody to channel their
instrumentation suggestions and ideas to this instrumentation working
group.  We'll then get a better understanding of what the user
community wants us to do in the future.  The current chairman of the
Instrumentation Working Group (IWG) is Richard McMahon at the
Institute of Astronomy and Nic Walton is the Secretary. [{Eds.\ Note:
more information can be found at the link to the ING Instrumentation
Development web pages at
http://www.ing.iac.es/Astronomy/astronomy.htm}]

{\em Roger Davies:} I thought that the reason that you were thinking
of standing down LDSS-2 was that you couldn't afford to operate it, not
that it was a candidate, necessarily, for new instrumentation money.

{\em Ren\'{e} Rutten:} The operations cost is one thing. Both LDSS-2 and
TAURUS-2 are in need of a major overhaul, simply because they are now
mature instruments. The AAO has upgraded its LDSS-1 to LDSS++, making
it more competitive.

{\em Roger Davies:} Can the observatory support LDSS-2, even in its
current form, to keep it running as it is.

{\em Ren\'{e} Rutten:} Everything can be done, but at a cost.

{\em Roger Davies:} In answer to your question, {\em your priorities for
new instrumentation should be INGRID and NAOMI}, they go
together. That's absolutely vital. That would be my personal opinion.

But I was talking about a different thing before. I have heard several
times that LDSS-2 might be stood down because of the operating
problems, so I was very keen that it shouldn't be stood down. I was
quite surprised by your numbers, because your numbers can be
summarised as: 40\% ISIS, 20\% prime focus, 20\% collimating
beams---LDSS-2 and TAURUS-2, and 20\% high resolution.  It is not clear
which of those 20\%s you would stand down from a scientific point of
view. 

Here you are doing it on the basis that the instruments in the
collimated beam area need more technical attention.  That is a perhaps
a {\em dangerous way to prioritise}, the scientific priorities might
suggest another set of instruments for de-commissioning.  I would say
that LDSS is a very important scientific instrument for a lot of
people, and maybe, at least should be kept going.

But in the area of new instrumentation I would absolutely
agree that you {\em have to get INGRID and NAOMI into a world
competitive state in the next year}, and that would be my priority
ahead of LDSS.

{\em Ren\'{e} Rutten:} My fear with LDSS is that if we just keep it going,
it will slowly become less reliable, less competitive. Users will get
less science from it, and eventually it will no longer be in demand.
Lets decide now that we don't want to use it anymore.

{\em Phil Charles:} Is it scientifically essential to have LDSS and
TAURUS in both hemispheres?

{\em Richard Bower:} We are using the wide-field camera to do surveys
to find distant clusters.  We want to be able to follow them
up. The nature of ROSAT pointings are that they are at high latitude
so they simply can not be followed up from the south. 

{\em Ray Sharples:} It is not the only area that is duplicated. There
is the clone of UES [{\em the Utrecht Echelle Spectrograph on the
WHT}] at the AAT. There is a proposal for an 8k mosaic at the AAT. I
think the argument that this is the only duplicated area is not
appropriate at all.

{\em Paul Groot:} Could you do your follow-ups with an instrument like
AF2+WYFFOS if you used two pointing instead of one to get round the
cluster crowding problems when placing fibres?

{\em Richard Bower:} It's possible in five pointings, with half an
hour dead-time between fields, thus with hour or two hour exposures
you have large overheads.

{\em Steve Smartt:} One of the arguments against LDSS on the WHT is
that GMOS with Gemini will go a lot deeper, thus LDSS will not be
competitive. 

{\em Roger Davies:} You should not be using 8-m telescopes to take
spectra of twenty-first and twenty-second magnitude galaxies. Thus
LDSS will still have a role.

{\em Nic Walton:} Some 8-m instruments have a large field, with high
multiplex. VIRMOS/NIRMOS on the VLT will be able to observe hundreds
of objects at a time, and with their high dynamic ranges, could
observe the bright 'LDSS' type objects as well. 

{\em Roger Davies:} The field of view on 4-m telescopes is linearly
twice that off 8-m telescopes.  You can do four with Gemini, or
one pointing with LDSS. You get deeper in the same
integration time with Gemini but it takes four pointings.

{\em Nic Walton:} OK, but with the upgrade to LDSS++ at the AAT, the
multiplex has increased by a factor of ten. If LDSS isn't given this
type of upgrade it becomes uncompetitive, even against similar
instruments on a 4-m.

{\em Phil Charles:} This comes down to the resources. As Ren\'{e}
said, we are going to have to put the resources either in the manpower
or face the consequences in terms of the implication for what Roger
has admitted ought be the top priority. There is only so
much that we can do with the resources available now.

{\em Reynier Peletier:} Is the prime focus really essential on the WHT,
because over half the INT time is now devoted to wide field imaging.

{\em Phil Charles:} What do people feel about dispensing with WHT prime
focus imaging, user facility.

{\em Clive Tadhunter:} This was discussed on the GBFC and I think it
was felt there that really we are too behind the competition.  There
are so many other similar facilities elsewhere on 4-m telescopes that
it wasn't really worthwhile developing this for the WHT. 

{\em Simon Hodgkin:} What is the point of developing Cassegrain
infrared imaging on the WHT when it can be done so much better from
UKIRT for example.

{\em Ren\'{e} Rutten:} The INGRID development, the camera, is essential for
the adaptive optics system.  Thus it can be used, as a no-cost extra,
for direct imaging at the Cassegrain focus.

{\em Nic Walton:} AF2+WYFFOS could be developed to give a J+H near IR
capability at low cost. Extending this further into the thermal K
region would require significant investment. Is there any desire for
this in the community?

{\em Jim Emerson:} Isn't it the case that the fibres don't transmit
much at K anyway?

{\em Nic Walton:} We wouldn't use fibres with K, probably an image
slicer design feeding a spectrograph at the 24 arcmin Cassegrain
focus. 

{\em Pat Roche:} We should point out that IRIS-II is coming on-line at
the AAT, sometime in the next year or two. Which gives you an eight
arc minute field at K using multi-slits.

{\em Nic Walton:} The WHT's Cassegrain field is significantly larger
than this. 

{\em Jim Emerson:} When we have limited resources we perhaps should not
be trying to do exactly the same thing on two different telescopes.
Getting into niches where other telescopes already are, be they UKIRT
or indeed the AAO, is very dangerous. Clearly there are two
hemispheres and may be very important to have similar facilities in
both, but INGRID and AO seem the thing to go for. You have to be very
careful not to add more things on when you are saying that you can not
run what you already have got.

{\em Nic Walton:} You can leverage your present capabilities by small
investments: AF2+WYFFOS for example can be enhanced in obvious
directions into the J+H bands via the upgrade to TEIFU\cite{sharples}
giving a significant new capability at low cost.  You have to think
about continuous renewal\cite{bohannan}, you can't stand still with
instrumentation, you must make incremental upgrades to give quite
significant performance improvements. Make use of new technological
breakthroughs to give big performance improvements at fairly low cost.

There is a much bigger question mark in going to K for spectroscopy on
the WHT --- that is major new investment that may not give a
significant capability when compared to Gemini.

{\em Paul Groot:} I think {\em you are right in not pursuing wide-field
imaging on the prime focus of the WHT} when you already have the
wide-field camera on the INT. Why offer the prime focus camera on the
WHT, de-commission it now and use the saved resources in other areas.

{\bf New Instrumentation: Superconducting Tunnel Junctions.} 

{\em Phil Charles:} Seeing that Tony Peacock is here, what sort of
time scale are we looking at before superconducting tunnel junctions
[{\em STJ's}]
\footnote{http://astro.estec.esa.nl/SA-general/Research/Stj/STJ\_main.html}
become a productive instrument?

{\em Tony Peacock:} I think we are talking about a 6$\times$6 pixel
array at the moment, we would be into 12$\times$8 by about the middle
of next year and we would have a 1000 element array ready by the third
quarter of next year. This would be running at 350 milli-kelvin with a
resolving power of about 10 at 500 nanometers, with a waveband 
coverage of 300 nanometer to two microns.

The other development which we are trying to do in parallel is a
technology development on the cooling system. We are linking up arrays
of tunnel junctions to a closed cycle mechanical system which will
mean that you will effectively only need a cooling procedure and
power, no consumables, and this would be running also at 300 milli
kelvin.

A Development which is slightly longer term is to improve the
resolving power albeit we would then have to reduce the temperature.
This would be on a time scale of the next three to five years. We are
aiming at a resolving power of 500 at 500 nanometers.

{\em Roger Davies:} Is that the limit?

{\em Tony Peacock:} No that's the limit imposed by our current cooling
capability.

{\em Roger Davies:} You have to get to a lower critical temperature
super conductor to get the resolution?

{\em Tony Peacock:} Yes, we can do that via an adiabatic demagnetiser
which goes through a mechanical cooler.  This would bring us down
to about something like five to ten milli kelvins.

{\em Roger Davies:} So then that resolution of 500 is as far as you can
go at that temperature.

{\em Tony Peacock:} Yes it is as far as we can go with our current
understanding of the filter film technology.

{\bf New Instrumentation: Adaptive optics and laser guide stars.}

{\em Pat Roche:} There are also lasers. This falls within AO and would
be a substantial investment.

{\em Phil Charles:} We've  heard that the La Palma site is as
good as Hawaii in terms of the spatial resolution as long as we are
working at H band and shorter.  We  want to look at the relative
priorities of this. 

{\em Vik Dhillon:} I would like to ask Andy a question\cite{longmore}:
the gains from an optical spectrograph with NAOMI seem amazing.  Would
you now say that's the highest priority instrument to develop for use
with NAOMI? Is this a higher priority than the IR spectrograph which I
think has been talked about by GBFC?

{\em Andy Longmore:} There are still some risks in that area. I think
there are niches I would feel fairly confident about, but we need to
see how NAOMI performs in the next few months. There are still some
things to resolve: you are not going to be able to get a broad wave
length coverage because the performance will fall off rapidly between
0.6 and 0.9 microns.  

You've got to look at optimising your instrument.  People should think
of instruments that have 50\% throughput with high resolution
spectroscopy, these will beat an 8-m below the J band.  These special
purpose instruments will give Ren\'{e} [Rutten] an extra problem because
you'll need a few of them to service a wide user base.  You may need
to optimise the throughput at one micron and you want something else
that's got a throughput at 0.6 microns, thus you may need two
instruments.

These were just a few ideas I had for specialist instruments, some of
which have been covered, for example single IFUs [{\em Integral Field
Unit}].  I think it is very important to pursue to the Lincoln Labs
high efficiency CCDs because that is exactly where you are going to
make a gain in the one micron band.

Small spectrometers --- if you're working with a 0.1 arc sec slit and
everything else is equal, then you can have a grating or an instrument
that is a fifth the size of people working with a half arc second
slit. You could have a series of small optimised spectrometers.
Coronographs need to be very well optimised for AO systems to get the
most out of them. CIAO on Subaru [{\em the Japanese 8-m
telescope}]\footnote{http://www.noaj.org}, for example, has rotating
pupils to take into account the rotation of the pupil on an alt-az
system.

Other areas to develop: infra-red wavelength sensors for tip-tilt
corrections will help when observing in dark clouds.  Higher order
correction, using new wavefront sensor arrays (faster, lower
readout, infrared) will give better AO performance in some circumstances.

Finally, you always may get some irregularities at the 1\%, 2\%, level
in your point spread functions.  There are ideas in terms of twin
channel functions which split the wavelength to two very close
neighbouring wavelengths that should have different spectral
properties.

{\bf New Instrumentation: Tip-Tilt Systems.}

{\em Reynier Peletier:} Would a simple tip-tilt correction system,
such as the tip-tilt secondary at UKIRT which really has improved
their images, that be an affordable  improvement for the WHT?

{\em Tim Hawarden:} The improvement that results is not necessarily
because we are doing adaptive optics. Some of the evidence suggests
that we are simply taking out telescope motions at UKIRT.

{\em Ren\'{e} Rutten:} In the optical, the gains from just tip-tilt
corrections are very limited.  You might get slightly sharper images
but the point spread function has a very broad wing and particularly
in the optical that is very bad.

{\em Tim Hawarden:} You should not underestimate the degree to which
your telescope shakes. The CFHT thought they had a solid telescope but
when they switched on HRCAM [{\em a high resolution imaging camera}]
they found strong evidence for telescope shake. Every telescope should
have a tip-tilt secondary because there is a lot of gain to be
made. The vibrations that you don't know about are still affecting the
optical quality of the images.  Our seeing on UKIRT, in terms of the
general average seeing, is better than is reported at the CFHT, so
CFHT probably has some dome seeing that we don't. But we have a
tip-tilt system and the other telescopes don't.

{\em Johan Knapen:} A potential problem with AO is that you can not
observe any object with AO because most objects would not have a
suitable point [to correct on] source either nearby or within them. 

{\em Bruce Bohannan:} It's not the seeing that we want, it's the
delivered image quality: it's the size of the image that is on the
detector so if you took a 30 second exposure and its the same as a
five minute exposure you don't need tip-tilt. But if your seeing is
significantly increased over that, then a tip-tilt secondary would
give you large gains in spectroscopy particularly.

{\em Richard Myers:} We've got many power spectra of image motions on
the WHT over the years. In general the gains you could get from full
aperture tip-tilt in the optical are simply negligible.  If you look
at short exposure images on the WHT, say very short millisecond
exposures, and do shift analysis to simulate what a perfect noise free
tip-tilt systems would give you providing it is not very windy, there
really are very negligible gains.

{\em Reynier Peletier:} It could be an experiment and it could be very
cheap and you could potentially gain a lot even though you think that
you know everything.  ESO's new understanding of their 3.6-m telescope
shows that we do not always know how the telescope operates and
behaves.

{\em Tim Hawarden:} The real test is, if you do millisecond exposures
and then co-add a bunch of them and your image is just as small after
you have co-added 10,000 of them, then you know tip-tilt isn't going
to improve things. 

{\em Roger Davies:} Ren\'{e} [Rutten] showed us this morning that PPARC is
looking at buying into the Spanish GranTeCan at the 10\% level by
making some changes in the way ING operates, transferring resources
from its development and operational lines to the GranTeCan.  I think
this group should at least consider whether 10\% is an adequate amount
to service our needs and whether 10\% is worth the cuts to the ING
that are proposed.

{\em Phil Charles:} This is why we made our JIF [{\em UK's Joint
Infrastructure Fund}] bid to join GranTeCan
at the 30\% level because we recognise that we don't have enough 8-m
telescope time.  The 10\% is the most we can afford if we get no more
money, if we are unsuccessful with JIF, but exploit the investment we
have in La Palma. We would need to make changes to our operation of
the ING so that we can make a major contribution into GranTeCan.
[{\em Ed's note: subsequently to this meeting it was announced that
the bid to JIF to join GranTeCan at the 30\% level failed. The bid to
fund VISTA, the UK's Visible and Infra-Red Survey Telescope for
Astronomy was approved.}]

{\em Tim Hawarden:} What is the 10\% going to cost us. How many nights
of 4-m time do we have to give to get one night of 8-m time.

{\em Ren\'{e} Rutten:} The details depend on the outcome of the
negotiations between the Netherlands, UK and Spain.  Broadly we are
discussing a redirection of $\pounds0.7-1$ million per year over the
period of ten years. This is released from the ING budget by reducing
the development programme by a factor of two and not introducing the
financial cut of 10\% after 2002 that PPARC has presently planned for
the ING.  Staff effort currently on La Palma which supports and
develops the ING telescopes, will be partly channeled into GranTeCan.
This means less enhancements and a lower level of development for the
ING.

{\em Phil Charles:} The negotiations are ongoing, I am involved in
these.  We need to exploit funding opportunities outside of the PPARC
area which may give us a chance to contribute to a new sea-level base
to be run jointly with the GranTeCan operation on La Palma.  That
would help us operate more efficiently.  These ideas indicate we can
achieve a 10\% involvement without any real additional funding.

[{\em Eds.\ note: We have edited out the discussion of operation funding
which ensued on the basis that the bid to buy into GranTeCan would
succeed.}]

{\em Colin Vincent:} There are a number of other bids apart form the
GranTeCan which equally may be successful and they are all only
capital bids. Hence PPARC will have to find the running costs if any
are successful. It is not really true to say that the problem is
wholly one of the ING.  The GBFC will look at the whole picture if one
of these bids is successful and see where the operating costs will
come from.

{\em Wilfred Boland:} How attractive is the ING when you reduce the
development budget by 50\%.

{\em Phil Charles:} In the mid-term we have to consider a WHT which
will not be operating in the work horse mode that it is now.  The
4-m's will work differently in a 8-m era.

{\em Colin Vincent:} The impact would be that whatever you plan to
develop would be delivered on a slower time scale. It does not mean to
say that you can not deliver large instruments but that they would be
delivered more slowly.  If you look at other telescopes, a lot of them
haven't got a long term development wedge as such, they bid on a project
by project basis.  The ING must bid for the resources it needs for its
highest priority projects.

{\bf La Palma and GranTeCan} 

{\em Phil Charles:} What are the relative priorities of GranTeCan, or
trying to get involved with SALT [{\em the South African Large
Telescope}]\footnote{http://www.salt.ac.za}.

{\em Jim Emerson:} A comment about the mode of getting money.  For a
long time in the UK we have imagined that the only source of money is
PPARC.  Recently various people have got money from other sources. The
Liverpool group have got money for their robotic telescope from the
European community and various other places. Maybe there is some money
from the JIF.  People ought to explore alternative funding routes as I
do not see how all the various astronomy proposals can be funded by
PPARC. In this context Public Understanding of Science activities are
very important in raising the profile of astronomy amongst potential
financial donors.

{\em Peter Sarre:} Possible participation in SALT seems to be
extraordinary good value for money both in terms Capital Investment
and ongoing costs.  It doesn't involve travel costs and seems to
guarantee continued UK access to the SAAO's [{\em South African
Astronomical Observatory}]\footnote{http://www.saao.ac.za} 1.9-m
telescope.

{\em Phil Charles:} We are almost at the time we thought we would
finish.  We haven't really talked very much about whether you feel sub
8-m telescopes have a future. I obviously feel very strongly that they
do. And I think that is a very important case to get across to the
funding agencies.

{\em Matt Burleigh:} You did not mention\cite{charles} that there is no
capability on the ING telescopes to do {\em fast}  photometry, a
capability which has completely disappeared from the ING telescopes.

{\em Phil Charles:} We have got half a dozen major groups in the UK
that want to do fast photometry. The
ULTRACAM\footnote{http://wwww.shef.ac.uk/~phys/people/vdhillon/ultracam/}
instrument will fill that role\cite{dhillon}.

{\em Nic Walton:} High speed photometry will be re-introduced in early
2000 as a by product of the ING's new new data acquisition
system. This should give times resolution of 10 to 20 milli-secs. 

{\em Ron Hilditch:} In the context of science subject orientated
planning of the use of the telescopes I think it imperative that the
JKT [{\em the 1.0-m Jacobus Kapteyn Telescope}] is maintained in its
current operational state, preferably with a high speed
photometer. The JKT is need to support very high resolution
spectroscopy for which we need UES on the WHT, if you take one of
those things away the whole programme is compromised.

{\em Phil Charles:} Ren\'{e} was also telling me about the new data
acquisition system.  The advantage of the SAAO high speed photometry
system is that you actually have zero dead time, and that may still
make those kinds of CCDs advantageous over even just an upgraded data
acquisition system.

{\em Paul Groot:} The WHT is very flexible in its ability to change
instrumentation and respond to targets of opportunity. It is also
excellent as a test bed for new technologies such as the STJ's.  I
think that is a very strong point for the future of the telescopes
like the William Herschel.  To give one example where a fast response
paid off is the discovery of the first optical counter part to a gamma
ray bust, observed at the WHT.  I think it is important not to loose
this flexibility when changing the way of operating these telescopes.

{\em Ren\'{e} Rutten:} I would like to thank everybody for the input, and
let the input from the user community not stop here.  Let us hear what
you want out of the ING telescopes, in the near future and in the
distant future, speak to us and let the Instrumentation Working Group
know what your needs are so that we can incorporate them into our
plans.

{\em Phil Charles:} Thank you very much everybody for your
contributions this afternoon.  Enjoy the rest of the conference.

\ack

The editors would like to thank Carmen Inarres, Jane Metz and Mark
Hughes, postgraduate astronomy students at the University of
Sheffield, for their help in recording the discussion session.


\begin{thebibliography}{999}

\bibitem{bohannan} Bohannan, B. 2000, NewAR, these proceedings.

\bibitem{bailey}  Bailey, J. 2000, NewAR, these proceedings.

\bibitem{charles} Charles, P.A. 2000, NewAR, these proceedings.

\bibitem{dhillon} Dhillon, V.S. 2000, NewAR, these proceedings.

\bibitem{knapen} Knapen, J.H.. 2000, NewAR, these proceedings.

\bibitem{longmore} Longmore, A.J. 2000, NewAR, these proceedings.

\bibitem{roche}  Roche, P. 2000, NewAR, these proceedings.

\bibitem{rutten}  Rutten, P. 2000, NewAR, these proceedings.

\bibitem{sharples} Sharples, R. 2000, NewAR, these proceedings.

\end{thebibliography}
\end{document}